\documentclass[12pt]{article}
\pdfoutput=1

\usepackage{scicite}
\usepackage{graphicx, siunitx}

\usepackage{times}

\newenvironment{sciabstract}{%
\begin{quote} \bf}
{\end{quote}}

\title{Nonlinear magnon control of atomic spin defects in scalable quantum devices} 

\author
{Mauricio Bejarano,$^{1,2}$ Francisco J. T. Goncalves,$^{1}$ Toni Hache,$^{3}$ Michael\\
Hollenbach,$^{1,4}$ Christopher Heins,$^{1}$ Tobias Hula,$^{1,5}$ Lukas Körber,$^{1,4}$ Jakob\\ 
Heinze,$^{1}$ Yonder Berencén,$^{1}$ Manfred Helm,$^{1,4}$ Jürgen Fassbender,$^{1,4}$ Georgy\\
V. Astakhov,$^{1}$ Helmut Schultheiss$^{1\ast}$\\
\\
\normalsize{$^{1}$Helmholtz-Zentrum Dresden-Rossendorf, Institute for Ion Beam Physics and Materials}\\
\normalsize{Research, 01328 Dresden, Germany}\\
\normalsize{$^{2}$Faculty of Electrical and Computer Engineering, Technical University of Dresden,}\\
\normalsize{01062 Dresden, Germany}\\
\normalsize{$^{3}$Max Planck Institute for Solid State Research, 70569 Stuttgart, Germany}\\
\normalsize{$^{4}$Faculty of Physics, Technical University of Dresden, 01062 Dresden, Germany}\\
\normalsize{$^{5}$Institute of Physics, Technical University of Chemnitz, 09107 Chemnitz, Germany}\\
\\
\normalsize{$^\ast$To whom correspondence should be addressed; E-mail:  h.schultheiss@hzdr.de}
}

\date{}

\begin{document} 


\baselineskip24pt


\maketitle


\begin{sciabstract}
	Ongoing efforts in quantum engineering have recently focused on integrating magnonics into hybrid quantum architectures for novel functionalities. While hybrid magnon-quantum spin systems have been demonstrated with nitrogen-vacancy (NV) centers in diamond, they have remained elusive on the technologically promising silicon carbide (SiC) platform mainly due to difficulties in finding a resonance overlap between the magnonic system and the spin centers. Here we circumvent this challenge by harnessing nonlinear magnon scattering processes in a magnetic vortex to access magnon modes that overlap in frequency with silicon-vacancy ($\textrm{V}_{\mathrm{Si}}$) spin transitions in SiC. Our results offer a route to develop hybrid systems that benefit from marrying the rich nonlinear dynamics of magnons with the advantageous properties of SiC for scalable quantum technologies.
\end{sciabstract}


\section*{Introduction}

Atomic-scale crystallographic defects in wide-bandgap semiconductors are being explored as quantum bits, or qubits, due to their convenient characteristics for quantum information processing and computing \cite{QIPC_review,quantum_guidelines,QSD_computing,QSD_computing2}. These defects have highly localized electronic states isolated within the bandgap of the host material, which endows them with long spin coherence times at room temperature \cite{coherence1,coherence2,coherence3,coherence4,coherence5,SiC_D3}. Moreover, by having optically addressable non-zero spin states, these defects provide an optical interface for spin initialization and readout, which makes them attractive for both quantum sensing and computing applications \cite{rev_Castelleto,QSD_computing,QSD_computing2,review_sensing,kraus,mapping}. However, achieving scalable quantum technologies requires embedding the quantum system in an heterogeneous architecture, with disparate but complementing building blocks. For this purpose, the nascent field of quantum engineering has been exploring hybrid quantum architectures which involve, in particular, magnons (spin waves), the collective excitation of spins in magnetically ordered materials \cite{hybrid1,hybrid2,Toeno1,Andrich,IEEE_awschalom,hybrid_Axel,candido}. Hybrid systems consisting of quantum spin defects (QSD) and magnons benefit from large coupling strengths in the sub-gigahertz regime owed to the high spin densities in the magnon-hosting element \cite{IEEE_awschalom,hybrid_Axel,coupling1, coupling2,coupling3}, enabling energy-efficient quantum operation and transduction. Furthermore, due to the ease of fabrication and miniaturization, magnonic devices can be integrated into wafer-scale quantum circuits \cite{hybrid_Axel} for entangling distant qubits \cite{Andrich,trifunovic,kikuchi} and for locally increasing the microwave sensitivity in sensing applications \cite{Andrich,amplification,toeno2}. Localized magnon auto-oscillations point to full electrical control of spatially-separated spin qubits \cite{agility_SHNO, bipolar_SHNO}. More generally, by introducing magnons to the quantum architecture, a wide range of interactions and control tools can be accessed, adding more functionalities to the quantum system.

Despite the increased interest in developing hybrid magnon-QSD systems, most of the efforts have focused on utilizing nitrogen-vacancy (NV) defects in diamond as the spin centers and yttrium iron garnet (YIG) as the magnon-hosting material \cite{IEEE_awschalom,hybrid_Axel}. This poses several challenges for technological implementations as diamond and YIG are not easily integrated into wafer-scale quantum devices \cite{rev_Castelleto,IEEE_awschalom,hybrid_Axel}. To this end, recent efforts have been made to implement hybrid architectures in alternative QSD-hosting materials. Silicon carbide (SiC) as a spin defect platform stands out as it is a more technologically mature material, with established wafer-scale fabrication protocols that can be leveraged for on-chip quantum devices \cite{kimoto_cooper_2014,SiC_fab1,rev_Castelleto,IEEE_heremans,Riedel_VSi_qbits,acoustic}. Additionally, owing to its rich polytypism, SiC possesses a diverse number of known spin defects with similar optical properties as the NV center in diamond \cite{coherence_SiC,coherence_SiC2,coherence2,coherence4,falk,SiC_D3}. Regardless of its technological interest, a hybrid magnon-QSD system on SiC has remained elusive mainly due to a lack of a frequency overlap between the resonant magnons in the ferromagnetic material and the electron spin resonance (ESR) of the spin defect. For example, for the negatively charged silicon vacancy at the quasicubic lattice site in the 4H-SiC polytype, the $\textrm{V}_{\mathrm{Si}}$(V2), the zero-field ESR lies at 70~MHz \cite{kraus,V2_ZFS}, which is situated within the magnon bandgap of the most common ferromagnetic materials. In the high field regime, the ESR of the $\textrm{V}_{\mathrm{Si}}$(V2) never crosses the magnon spectrum due to having similar gyromagnetic ratios \cite{SiC_D3}.

In this work, we propose a hybrid magnon-QSD system that utilizes nonlinear magnon dynamics in a magnetic vortex to achieve magnon coupling to $\textrm{V}_{\mathrm{Si}}$ ensembles in 4H-SiC. The frequencies of the magnons involved are field-independent over a large magnetic field range \cite{lukas_thesis}, which facilitates the overlap with the $\textrm{V}_{\mathrm{Si}}$ spin transitions. We experimentally demonstrate a frequency downconversion of the applied microwave excitation which allows a pure-magnonic control of the spin qubit at previously inaccessible frequencies. Our results open new avenues for developing scalable hybrid quantum technologies while offering a testbed for exploring the convergence of nonlinear, threshold-activated systems with intrinsically linear quantum systems.


\section*{Results}

\subsection*{Hybrid magnon-QSD system}

\begin{figure*}[h!]
\centering
\includegraphics[width=1.0\textwidth]{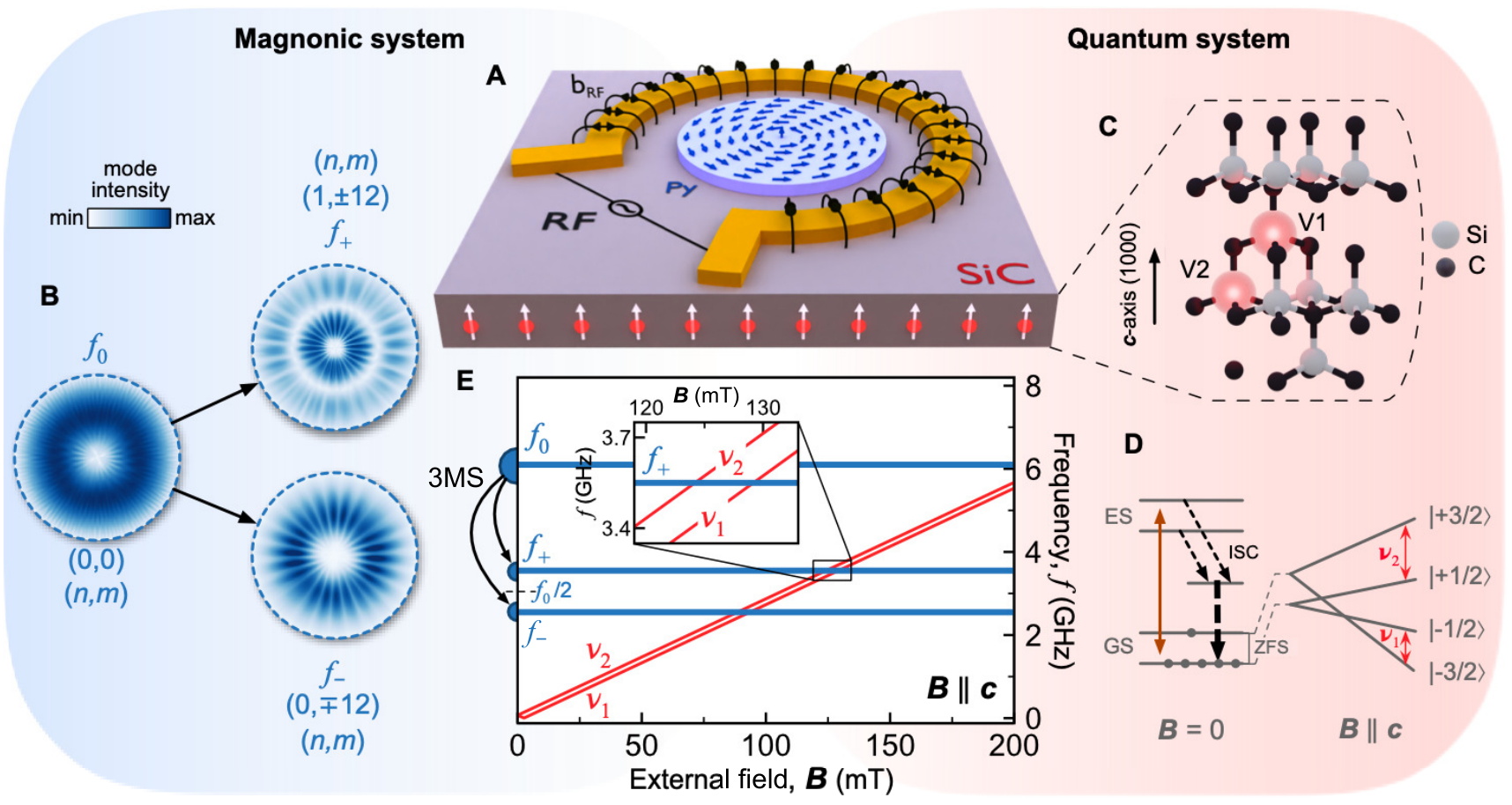}
\caption{\textbf{Proposed hybrid magnon-quantum spin defect system in SiC.} (A) A \SI{5.1}{\micro\meter}-diameter permalloy (Py) disc in a vortex state configuration lies on top of a SiC substrate hosting $\textrm{V}_{\mathrm{Si}}$ ensembles. The magnetization dynamics of the disc are excited by oscillating magnetic fields (black arrows) from an on-chip antenna. (B) Magnons inside the disc, described by their radial and azimuthal indices $(n,m)$, undergo three-magnon scattering (3MS) processes when subjected to high-power AC magnetic fields. The initial radial magnon mode $f_0$ scatters into azimuthal modes $f_+$ and $f_-$ following specific selection rules. (C) Crystallographic structure of 4H-SiC showing the two inequivalent lattice sites for $\textrm{V}_{\mathrm{Si}}$, namely V1 and V2. (D) Optical cycle of $\textrm{V}_{\mathrm{Si}}$(V2) and Zeeman splitting of its ground state energy levels at $B \gg 2.5$ mT. The two most probable spin transitions, $\nu_{1}$ and $\nu_{2}$, are indicated by arrows. (E) Frequency overlap between the ESR of the $\textrm{V}_{\mathrm{Si}}$(V2) defect in SiC and the scattered magnon modes within the disc. Inset shows a zoom into the crossing between $f_+$ and the $\nu_{1}$ and $\nu_{2}$ resonances of $\textrm{V}_{\mathrm{Si}}$(V2). The spatial profiles of the modes in (B) and their resonances in (E) were obtained with micromagnetic simulations. }\label{fig1}
\end{figure*}

Our proposed hybrid magnon-QSD system, shown in Fig.~\ref{fig1}, has two main components: a magnonic subsystem serving as an on-chip microwave transducer and a quantum subsystem with an ensemble of spin defects to be controlled. The magnonic subsystem consists of a disc of $\textrm{Ni}_{81}\textrm{Fe}_{19}$, also known as permalloy (Py), with \SI{5.1}{\micro\meter} in diameter and \SI{50}{\nano\meter} in thickness in a vortex state configuration (Fig.~\ref{fig1}A). The confined magnon eigenmodes of the vortex state disc (Fig.~\ref{fig1}B), which have been extensively studied before \cite{vortex1,vortex2,vortex3}, are characterized by their indices $(n,m)$ with $n=0,1,2,...$ determined by the number of nodes in the radial direction and $m=0,\pm1,\pm2,...$ determined by the number of periods in the azimuthal direction. These modes appear in the magnon spectrum as degenerate duplets with opposite $m$. To excite these vortex magnons we utilize an on-chip omega-shaped antenna surrounding the disc (see Fig.~\ref{fig1}A) which yields predominantly out-of-plane oscillating magnetic fields at the center of the antenna \cite{supp}. Due to the rotational symmetry of this excitation field, only magnons with $m=0$ can be directly excited by the antenna.

The quantum subsystem (Fig.~\ref{fig1}C,D) is composed of a $\textrm{V}_{\mathrm{Si}}$ layer in 4H-SiC created by proton irradiation and localized approximately \SI{200}{\nano\meter} below the surface \cite{supp}. As shown in Fig.~\ref{fig1}C, two types of $\textrm{V}_{\mathrm{Si}}$ defects at inequivalent lattice sites can be created \cite{Janzen_V1V2}. Fig.~\ref{fig1}D shows the energy levels and optical transitions of the $\textrm{V}_{\mathrm{Si}}$(V2) defect, which is the one we focus on in this work. The energy levels of this defect are quadruplets with $S=3/2$, split at both ground and excited states into ${m_s = \pm1/2,\pm3/2}$ \cite{SiC_D,SpinVSi,SiC_D3,review_Georgy}. A spin-dependent intersystem crossing (ISC) enables spin polarization and readout via photoluminescence (PL). At zero external field, the energy levels are spin-split by the zero-field splitting (ZFS) of $2D =$~\SI{70}{\mega\hertz}, where $D$ is the crystal field constant. In the high field regime, at $B \gg$~\SI{2.5}{\milli\tesla}, the $\textrm{V}_{\mathrm{Si}}$ spin resonance frequencies shift linearly with $B$ due to Zeeman splitting following 
\begin{equation}
\nu_{1,2} =  \gamma B \pm D(3\cos^{2}\theta -1)
\end{equation}
with ${| \gamma |=}$~\SI{28}{\mega\hertz}$/\mathrm{mT}$ the electron gyromagnetic ratio, and $\theta$ the angle between the external magnetic field and the $c$-axis \cite{SiC_D,SiC_D2,SiC_D3}. The resulting $\textrm{V}_{\mathrm{Si}}$ spin state is interrogated via optically-detected magnetic resonance (ODMR) \cite{kraus}.

In order to obtain a frequency crossing between the two subsystems we utilize nonlinear magnon scattering events taking place within the disc. When excited above a certain microwave-power threshold, the antenna-driven radial magnon mode ($m=0$) with frequency $f_0$ splits spontaneously into two secondary azimuthal magnon modes ($m\neq0$) with frequencies ${f_- = f_0/2 - \Delta f}$ and ${f_+ = f_0/2 + \Delta f}$ \cite{katrin,3MS_theory}. The selection rules of this three-magnon scattering (3MS) process require these secondary modes to have opposite azimuthal index $m$, but different radial indices $n$ which, in return, leads to a frequency split $\Delta f$ between them, as shown in Fig.~\ref{fig1}B. Via this frequency downconversion, these scattered modes cross the $\textrm{V}_{\mathrm{Si}}$ spin resonance frequencies at two distinct field regions as shown in Fig.~\ref{fig1}E. In this figure the horizontal lines are the frequencies of the magnon eigenmodes as obtained by micromagnetic simulations while the diagonal lines are the $\nu_{1}$ and $\nu_{2}$ spin resonances of the $\textrm{V}_{\mathrm{Si}}$(V2) defect obtained by analytical calculations using equation (1). Through 3MS, the radial mode $f_0$ excited at \SI{6.1}{\giga\hertz} splits into the azimuthal modes $f_-= $~\SI{2.55}{\giga\hertz} and $f_+= $~\SI{3.55}{\giga\hertz} enabling pure magnon-driven excitation of the $\textrm{V}_{\mathrm{Si}}$ defects at the crossing points $B \parallel c =$~\SI{99}{\milli\tesla} and ${B \parallel c =}$~\SI{127}{\milli\tesla} (Fig.~\ref{fig1}E). At these external magnetic fields, the direct antenna excitation and the dynamic dipolar fields from the $f_0$ mode are off-resonant to the $\textrm{V}_{\mathrm{Si}}$ spins.

\subsection*{Experimental characterization of the non-interacting conditions}

\begin{figure*}[h!]
\centering
\includegraphics[width=1.0\textwidth]{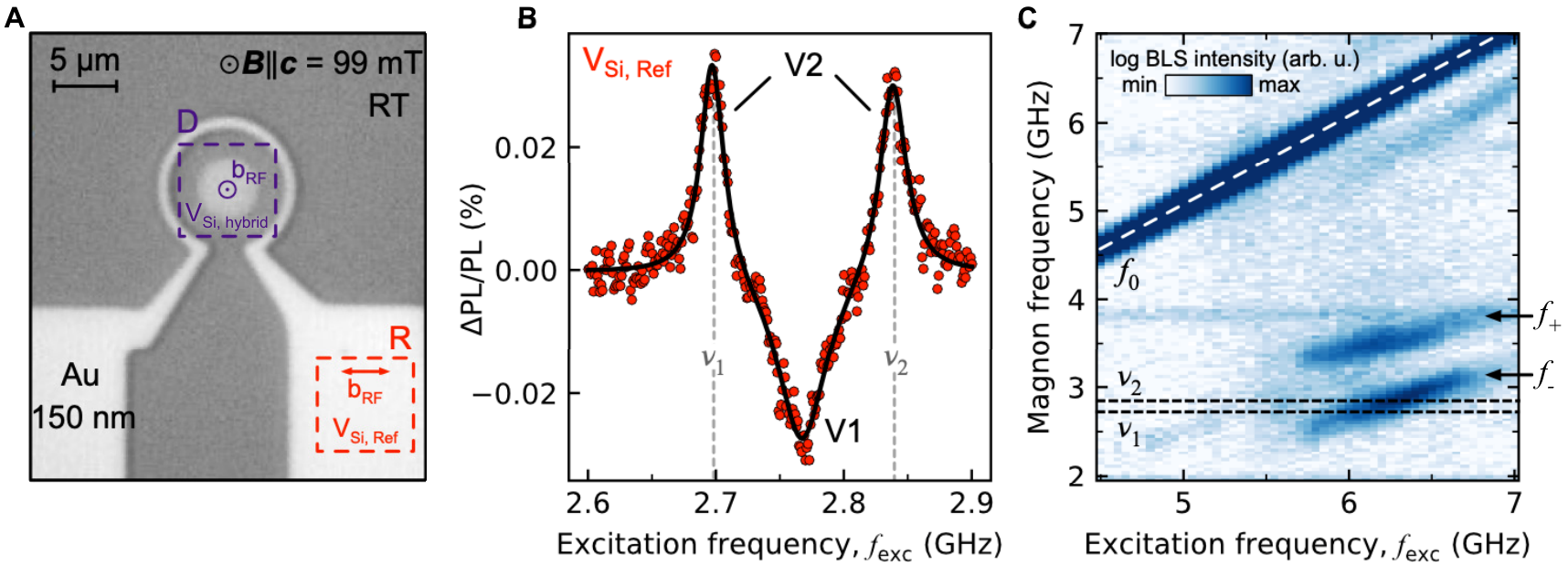}
\caption{\textbf{$\textrm{V}_{\mathrm{Si}}$ spin resonance and magnon modes at 99 mT.} (A) Optical microscope image of the antenna and the permalloy disc showing two distinct sites, D and R, measured through the backside of the SiC substrate. The magnons and the resonances of the $\textrm{V}_{\mathrm{Si}}$ defects below the disc, $\textrm{V}_{\mathrm{Si,hybrid}}$, were interrogated at site D. A reference $\textrm{V}_{\mathrm{Si,Ref}}$ ODMR spectrum was measured at site R. All the measurements were done at room temperature. (B) ODMR spectrum of $\textrm{V}_{\mathrm{Si,Ref}}$ with -3 dBm of microwave power. The peaks at $f_{\mathrm{exc}}=$~\SI{2.7}{\giga\hertz} and $f_{\mathrm{exc}}=$~\SI{2.84}{\giga\hertz} are attributed to the $\nu_{1}$ and $\nu_{2}$ spin transitions distinctive of the $\textrm{V}_{\mathrm{Si}}$(V2) defect. The negative contrast signal is associated with the $\textrm{V}_{\mathrm{Si}}$(V1) defect. The solid black line is a Lorentzian fit. (C) \textmu BLS spectrum of the disc at 21 dBm of microwave power showing the nonlinear splitting of the $f_0$ mode into the $f_-$ and $f_+$ modes. The horizontal dashed black lines show the $\nu_{1}$ and $\nu_{2}$ $\textrm{V}_{\mathrm{Si}}$(V2) resonances extracted from (B).}\label{fig2}
\end{figure*}

We now turn to analyze the resonant response of both the magnonic and quantum subsystems independently at the first frequency crossing point ($B \parallel c =$~\SI{99}{\milli\tesla}) as shown in Fig.~\ref{fig1}E. We identify two measurement sites, D and R, to separate antenna excitation from magnon excitation (Fig. \ref{fig2}A). The ODMR response of $\textrm{V}_{\mathrm{Si,Ref}}$ spins, located at measurement site R, is shown in Fig.~\ref{fig2}B. These defects can only couple to the oscillating magnetic fields from the antenna, as they are far away from the center of the disc. The positive contrast signal in Fig.~\ref{fig2}B corresponds to the PL fingerprint of the $\textrm{V}_{\mathrm{Si}}$(V2) defect (see Fig.~\ref{fig1}C) under high magnetic fields \cite{SiC_D2}. The two peaks stem from the $\nu_1$ and $\nu_2$ spin transitions (Fig.~\ref{fig1}D), separated by $4D = 140$ MHz, which indicates that the field is properly aligned with the $c$-axis ($\theta=0$). The negative contrast signal is attributed to the $\textrm{V}_{\mathrm{Si}}$(V1) defect \cite{diamond_prob1,supp}. 

For studying the nonlinear dynamics of our magnonic subsystem, room temperature micro-focused Brillouin light scattering microscopy (\textmu BLS) was used. Fig.~\ref{fig2}C shows the \textmu BLS spectrum at $B \parallel c =$~\SI{99}{\milli\tesla}, collected at measurement site D (Fig.~\ref{fig2}A), as a function of the microwave excitation frequency, $f_{\mathrm{exc}}$, at a fixed power above the reported nonlinear threshold \cite{katrin}. The white diagonal dashed line marks the linear magnon response, where the directly-excited mode follows the applied excitation frequency $f_0 = f_{\mathrm{exc}}$. At \SI{5.8}{\giga\hertz}~$<f_{\mathrm{exc}}<$~\SI{7}{\giga\hertz} two off-diagonal signals appear, which correspond to the scattered magnons $f_-$ and $f_+$, in accordance to the simulation data shown in Fig.~\ref{fig1}E. The black horizontal dashed lines highlight the spin resonances $\nu_1$ and $\nu_2$ of the $\textrm{V}_{\mathrm{Si}}$(V2) defect as extracted from Fig.~\ref{fig2}B. We can clearly see that there is a frequency crossing between both subsystems at $f_{\mathrm{exc}}\approx$~\SI{6.1}{\giga\hertz}, which matches the results shown in Fig.~\ref{fig1}E. At the first frequency crossing point ($B \parallel c =$~\SI{99}{\milli\tesla}) when ${f_{\mathrm{exc}}=f_0\approx}$~\SI{6.1}{\giga\hertz}, the scattered $f_-$ magnon mode will be resonant to the $\textrm{V}_{\mathrm{Si}}$ defects and, in turn, will drive spin transitions that induce changes in the PL emitted by the defects.

\subsection*{Threshold dynamics of the nonlinear excitation scheme}

\begin{figure*}[h!]
\centering
\includegraphics{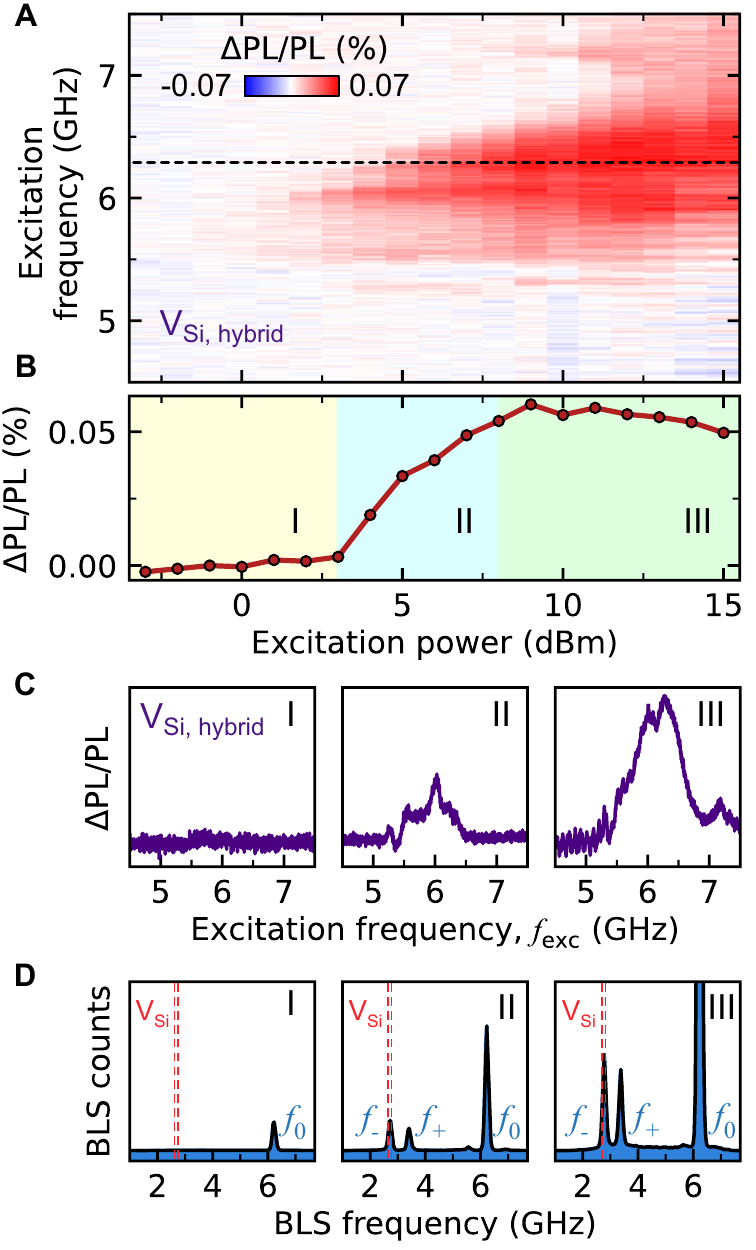}
\caption{\textbf{Threshold process of the nonlinear excitation scheme} (A) ODMR spectrum for increasing excitation powers at 99 mT measured at site D (see Fig.~\ref{fig2}A). (B) ODMR contrast integrated in a 20 MHz frequency window centered at 6.24 GHz (denoted by dashed line in (A)) showing the characteristic threshold behavior of 3MS processes. Three distinct regimes can be identified: I)-below threshold, II)-above thresold, and III)-saturation. For each regime, the average ODMR spectra are shown in (C) and the corresponding \textmu BLS spectra are shown in (D). }\label{fig3}
\end{figure*}

Utilizing the previous findings, we now experimentally corroborate the impact of the $f_-$ magnon on the $\textrm{V}_{\mathrm{Si}}$ spins in its proximity. Fig.~\ref{fig3}A shows the ODMR spectra of the $\textrm{V}_{\mathrm{Si,hybrid}}$ spins, located at site D directly below the disc (see Fig.~\ref{fig2}A), for increasing microwave excitation powers at the first frequency crossing point ($B \parallel c =$~\SI{99}{\milli\tesla}). As 3MS is a threshold process, a power dependence of the ODMR signal would reflect the nonlinear magnon population in the disc. It can be seen that at low microwave excitation powers there is no change in the PL from the $\textrm{V}_{\mathrm{Si,hybrid}}$ spins. However, at high microwave powers there is a strong $\Delta$PL/PL signal centered at $f_{\mathrm{exc}}\approx$~\SI{6.1}{\giga\hertz}. As discussed previously, at this external magnetic field the directly-excited mode $f_0 \approx$~\SI{6.1}{\giga\hertz} is off-resonant to the $\textrm{V}_{\mathrm{Si}}$ spins. This implies that the change in PL observed at $f_{\mathrm{exc}}\approx$~\SI{6.1}{\giga\hertz} can only be originated from a downconversion of the high frequency $f_0$ mode into the lower frequency $f_-$ mode which is resonant to the $\textrm{V}_{\mathrm{Si}}$ spins at $B\parallel c =$~\SI{99}{\milli\tesla} as shown in Fig.~\ref{fig2}C.

To elucidate the evolution of this nonlinear process, we extract the excitation-power dependence at $f_{\mathrm{exc}}=$~\SI{6.24}{\giga\hertz} and identify three separate regimes: I)-below threshold, II)-above threshold, and III)-saturation (Fig.~\ref{fig3}B). For each regime, Fig.~\ref{fig3}C,D show the average $\Delta$PL/PL of the $\textrm{V}_{\mathrm{Si,hybrid}}$ spins and the average magnon population intensity in the disc, respectively. Below threshold, there is no change in the PL signal coming from the $\textrm{V}_{\mathrm{Si,hybrid}}$ spins as there is no magnon overlapping the $\textrm{V}_{\mathrm{Si}}$ resonance frequency. In this regime, only the directly-excited magnon $f_0$, which is non-resonant to the $\textrm{V}_{\mathrm{Si,hybrid}}$ spins, is present. Above threshold, there is a non-zero \textmu BLS signal intensity at the frequencies of the scattered magnons $f_-$ and $f_+$ (Fig.~\ref{fig3}D). As the magnon $f_-$ matches the $\textrm{V}_{\mathrm{Si}}$ resonance frequency, it drives $\textrm{V}_{\mathrm{Si}}$ spin transitions which induce a change in PL. At saturation, the intensity of the scattered magnon $f_-$ is the largest, which gives rise to the high change in PL observed in Fig.~\ref{fig3}C. While further increasing the excitation power, higher-order interaction processes between the scattered magnons become relevant. These processes finally lead to decoherence between the directly-excited and the scattered magnons \cite{katrin,Lvov}. With that, the energy flux into the modes $f_+$ and $f_-$ is limited and a saturation in $\Delta$PL/PL is observed.

\subsection*{Field-dependent magnon excitation of $\textrm{V}_{\mathrm{Si}}$ defects}

\begin{figure*}[h!]
\centering
\includegraphics[width=1.0\textwidth]{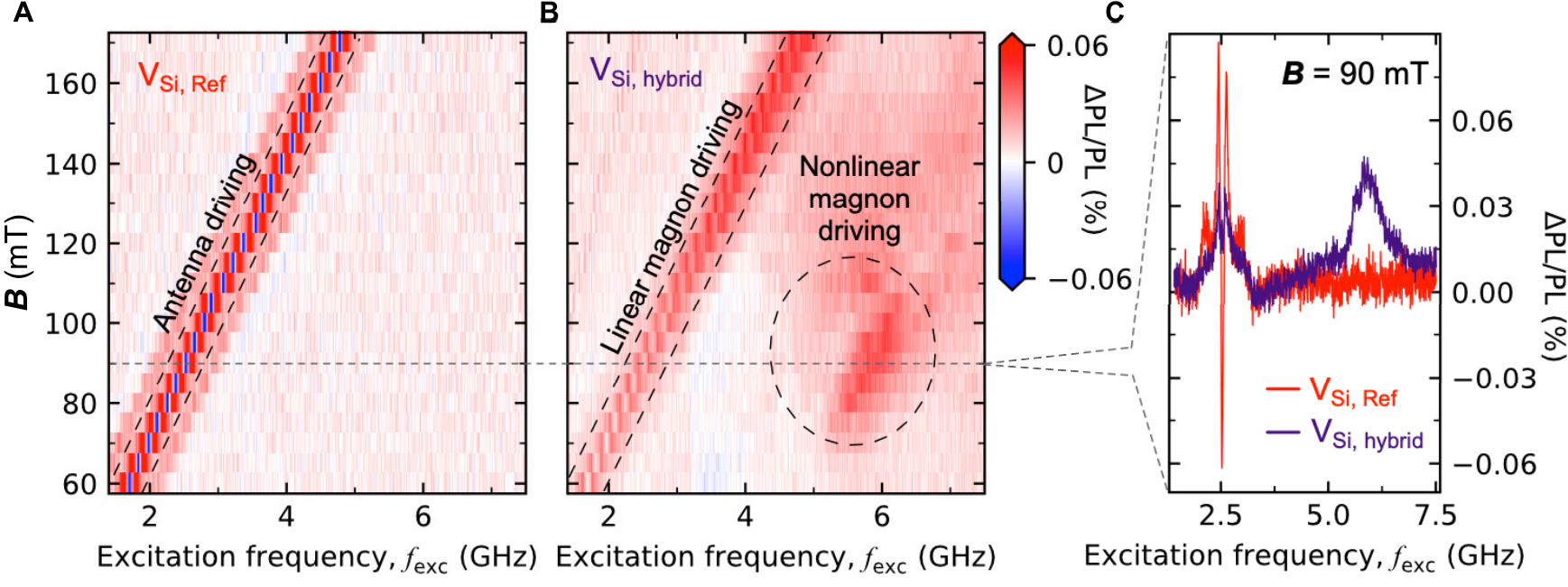}
\caption{\textbf{Nonlinear field-dependent excitation of $\textrm{V}_{\mathrm{Si}}$ defects.} (A) Evolution of the ODMR spectra from $\textrm{V}_{\mathrm{Si,Ref}}$ spins with increasing external magnetic fields,  measured at site R (see Fig.~\ref{fig2}A), i.e., far from the influence of the disc. (B) Magnetic field evolution of the ODMR spectra from $\textrm{V}_{\mathrm{Si,hybrid}}$ spins at site D, i.e., below the disc, showing linear and nonlinear magnonic excitation regimes. The off-diagonal signal around $f_{\mathrm{exc}}\approx$~\SI{6.1}{\giga\hertz} indicates the presence of the scattered magnon modes, which emit oscillating dipolar magnetic fields at the resonance frequency of the $\textrm{V}_{\mathrm{Si,hybrid}}$ spins. The dashed lines serve as guides. (C) shows line traces at 90~mT of the spectra in (A) and (B). The $\Delta$PL/PL peak at $f_{\mathrm{exc}}\approx$~\SI{6.1}{\giga\hertz} suggests the defects are being driven by the magnetic dipolar fields stemming from the scattered magnons in the disc. The spectra in (A) and (B) were both obtained using 9 dBm of microwave power which is above the 3MS threshold as inferred from Fig.~\ref{fig3}B.}\label{fig4}
\end{figure*}

Next, we move beyond the first frequency crossing point in Fig.~\ref{fig1}E and explore the frequency response of the $\textrm{V}_{\mathrm{Si,hybrid}}$ spins to the magnons in the disc over a wider magnetic field range ${B \parallel c}$. For comparison purposes, in Fig.~\ref{fig4}A we show a reference ODMR spectrum for the $\textrm{V}_{\mathrm{Si,Ref}}$ spins which are insensitive to the dipolar fields from the magnons in the disc (see Fig.~\ref{fig2}A). This spectrum shows the linear dependence of the $\textrm{V}_{\mathrm{Si}}$ spin resonance frequencies with respect to the field in the absence of the disc, which was discussed earlier and shown in Fig.~\ref{fig1}E. 

In Fig.~\ref{fig4}B we show the evolution of the ODMR spectrum for the $\textrm{V}_{\mathrm{Si,hybrid}}$ spins for increasing magnetic fields. Two distinct $\Delta$PL/PL signals can be identified: ($i$) A diagonal signal, linearly shifting with the field and ($ii$) an off-diagonal signal at \SI{5}{\giga\hertz}~$<f_{\mathrm{exc}}<$~\SI{7}{\giga\hertz} approximately. Despite similarities with the antenna-driven ODMR signal from the $\textrm{V}_{\mathrm{Si,Ref}}$ spins in Fig.~\ref{fig4}A, we attribute the linearly shifting signal in Fig.~\ref{fig4}B to the presence of the directly-excited $f_0$ magnon. On one side, at measurement location D, where the $\textrm{V}_{\mathrm{Si,hybrid}}$ spins are located, the antenna microwave magnetic fields cannot induce $\textrm{V}_{\mathrm{Si}}$ spin transitions given $\textrm{b}_{\mathrm{RF}} \parallel c$ \cite{supp}. As a result, the $\Delta$PL/PL signal in Fig.~\ref{fig4}B is entirely caused by the dynamic dipolar fields from the magnons in the disc. Furthermore, the low $\Delta$PL/PL signal intensity at low $f_{\mathrm{exc}}$ correlates to the off-resonant excitation of the $f_0$ mode \cite{katrin}. This mode, which reacts linearly with $f_{\mathrm{exc}}$, transduces the homogeneous out-of-plane antenna excitation $\textrm{b}_{\mathrm{RF}}$ to an in-plane microwave excitation which then drives changes in PL through $\textrm{V}_{\mathrm{Si}}$ spin transitions.

The off-diagonal $\Delta$PL/PL signal in Fig.~\ref{fig4}B can be explained by the 3MS process described before and shown in Fig.~\ref{fig1}B,E and Fig.~\ref{fig2}C. As exemplified with the line cuts in Fig.~\ref{fig4}C, at $B \parallel c=$~\SI{90}{\milli\tesla} the $\textrm{V}_{\mathrm{Si}}$ spin resonance frequencies are centered around $f_{\mathrm{exc}}\approx$~\SI{2.5}{\giga\hertz}. The peak in $\Delta$PL/PL signal at $f_{\mathrm{exc}}\approx$~\SI{6.1}{\giga\hertz} is then attributed to the spontaneous splitting of the directly-excited $f_0 =$~\SI{6.1}{\giga\hertz} magnon into the secondary magnon $f_-$ which, in turn, is resonant to the $\textrm{V}_{\mathrm{Si}}$ spins. It is important to note that for the $\textrm{V}_{\mathrm{Si}}$ spins to be resonant to $f_{\mathrm{exc}}\approx$~\SI{6.1}{\giga\hertz} the external magnetic field would have to be $B \parallel c=$~\SI{217}{\milli\tesla} which is not within the field range used for the experiments in Fig.~\ref{fig4}A,B.

\section*{Discussion}

We presented a hybrid magnon-QSD system with $\textrm{V}_{\mathrm{Si}}$ in SiC. Utilizing nonlinear magnon-scattering processes in a ferromagnetic disc in a vortex state, we excite magnon modes that are resonant to the $\textrm{V}_{\mathrm{Si}}$ spin defects. The resonant interaction between the magnon modes and the $\textrm{V}_{\mathrm{Si}}$ spins can be tuned by changing the external magnetic field as well as the microwave power being delivered by the lithographically patterned antenna which drives the nonlinear dynamics in the magnetic vortex. We envision that similar hybrid systems could be implemented with other spin defects in SiC, such as the divacancy complex (VV$^0$), which shows very promising quantum properties \cite{coherence_SiC2,IEEE_awschalom}. By implementing nonlocal stimulation of 3MS, controlling the $\textrm{V}_{\mathrm{Si}}$ spins could be achieved below the microwave power threshold \cite{nonlocal_lukas}. 

The excitation scheme proposed has two important implications: first, magnon excitation and antenna excitation are entirely decoupled. This enables pure antenna or pure magnon excitation, but not both at the same time. This is remarkably different from the commonly used approach of using magnons to mediate the control of spin defects in which the magnon excitation is added to the antenna excitation and the resulting microwave amplification is measured by the increased Rabi frequency \cite{Rabi1,Rabi2}. The second implication is that the excitation mechanism is nonlinear, due to the intrinsic nonlinearity of the underlying magnon scattering process. These two unique characteristics offer new pathways to explore the interaction between magnons and QSD, particularly important for developing magnetic field sensors with QSD and for controlling QSD with magnons either for fundamental or technological applications. Moreover, by bringing together nonlinear magnon physics and quantum systems, our system constitutes a proof-of-principle of a hardware platform to explore technologies, such as quantum neurons \cite{quantum_neuron} and quantum frequency mixers \cite{qt_mixers}, that benefit from the complementary advantages of both systems.



%

\section*{Acknowledgments}
We thank A. Henschke and R. Narkovic for their support with the experimental setup and with microwave simulations, respectively. We thank Ulrich Kentsch and the Ion Beam Center (IBC) at Helmholtz-Zentrum Dresden-Rossendorf (HZDR) for the proton implantation. Support by the Nanofabrication Facilities Rossendorf (NanoFaRo) at IBC is gratefully acknowledged. \textbf{Funding:} This work was supported by the German Research Foundation (DFG) under the grant SCHU 29224-1. \textbf{Author contributions:} Conceptualization: H.S., G.V.A. Experimental measurements: M.B., F.J.T.G., C.H. Device fabrication: T.Ha. Spin defect engineering: M.Ho., Y.B., G.V.A. Experimental setup development: M.B., F.J.T.G., J.H. Micromagnetic simulations and analytical calculations: M.B. Visualization: M.B., T.Hu., M.Ho., L.K., C.H. Supervision: H.S., G.V.A., F.J.T.G. Resources: M.He., J.F. Writing (original draft): M.B. Writing (review and editing): All authors contributed to manuscript preparation. \textbf{Competing interests:} The authors declare that they have no competing interests. \textbf{Data and materials availability:} all data is available in the manuscript or the supplementary materials.

\section*{Supplementary materials}
Materials and Methods\\
Supplementary Text\\
Figs. S1 to S4\\
References \textit{(59-60)}

\end{document}